\documentclass[twocolumn,aps,showpacs,pra,amsmath,amssymb,floatfix,superscriptaddress]{revtex4}
\usepackage{color}
\usepackage{graphicx}
\usepackage{dcolumn}
\usepackage{bm}
\usepackage{array}
\usepackage{float}
\usepackage{supertabular}
\usepackage{longtable}
\usepackage{mathrsfs}
\usepackage{txfonts}
\begin{document}

\title{Cooling through optimal control of quantum evolution}

\author{
Armin Rahmani}
\affiliation{
Theoretical Division, T-4 and CNLS, Los Alamos National Laboratory, Los Alamos, NM 87545, USA} 
   
\author{
Takuya Kitagawa}        
\affiliation{Physics Department, Harvard University, Cambridge, MA 02138, USA}  
              
\author{Eugene Demler}         
\affiliation{Physics Department, Harvard University, Cambridge, MA 02138, USA}             
            
\author{Claudio Chamon}
\affiliation{Physics Department, Boston University, Boston, MA 02215, USA}

\date{\today}

\begin{abstract}

Nonadiabatic unitary evolution with tailored time-dependent Hamiltonians can prepare systems of cold atomic gases with various desired properties such as low excess energies. For a system of two one-dimensional quasicondensates coupled with a time-varying tunneling amplitude, we show that the optimal protocol, for maximizing any figure of merit in a given time, is bang-bang, \textit{i.e.}, the coupling alternates between only two values through a sequence of sudden quenches. Minimizing the energy of one of the quasicondensates with such a nonadiabatic protocol, and then decoupling it at the end of the process, can result in effective cooling beyond the current state of the art. Our cooling method  can be potentially applied to arbitrary systems through an integration of the experiment with simulated annealing computations.

\end{abstract}

\maketitle

\section{Introduction}\label{intro}

Recent advances in the physics of ultracold atoms, brought about by cooling techniques such as evaporative and laser cooling, have stirred up great interest in the nonequilibrium dynamics of many-body quantum systems~\cite{Greiner2002,Kinoshita2006,Sadler2006,Hofferberth2007,Bloch2008}.  Creating the ground states of important model Hamiltonians, such as the two-dimensional fermionic Hubbard model, however, remains an outstanding challenge. Although such Hamiltonians can be created with cold atoms with a high degree of control over interactions and disorder and in relative thermal isolation, the effective temperatures one can reach with the current cooling techniques are still too high. Thus, expanding the boundaries of atomic cooling could open the door to the quantum simulation of unsolved condensed-matter models. In addition to cooling, preparing systems with other desired characteristics, such as, e.g., number-squeezed ones, is of considerable interest due to the potential applications to quantum metrology and precision measurements.

 Focusing on a a pair of two coupled elongated (assumed one-dimensional) quasicondensates (hereafter referred to simply as condensates despite lack of true long-range order) as an explicit exactly solvable example, we propose a scheme for preparing cold atomic systems with custom-ordered figures of merit through optimal control of their nonequilibrium quantum dynamics. As we will show, the large degree of dynamical control over these systems provides, among others, a new means of bringing them even closer to zero temperature.

Let us begin by giving a few examples of experimentally relevant quantities one can optimize in cold atom systems:
\begin{enumerate}
\item Effective cooling: by minimizing quantities such as the excess energy, number of quasiparticle excitations, or the trace distance between the density matrix of the system and its zero-temperature density matrix.
\item Phase coherence: by minimizing the fluctuations of the relative phase between two condensates (with spatially fluctuating phases), which is important for matter-wave interferometry~\cite{Andrews97, Polkovnikov06,Gritsev06,Jo07,Hofferberth2008}.
\item Number squeezing: by minimizing the particle number fluctuations of a system~\cite{Esteve08,Grond09, Julia-Diaz2012}, which is important, e.g., in precision measurements~\cite{Giovannetti04}.
\end{enumerate}
Focusing on effective cooling, we show in this paper that (at least) one of the condensates in the system of Fig. 1 can be cooled down by a factor of 5 with our proposed method under reasonable experimental conditions. This is not a fundamental bound, however, and cooling by several orders of magnitude is in principle possible for highly asymmetric systems. To cool a generic quantum system, we propose a scheme based on the integration of experimental measurements of excess energy and Monte-Carlo simulations.
\begin{figure}
 \includegraphics[width =.9\columnwidth]{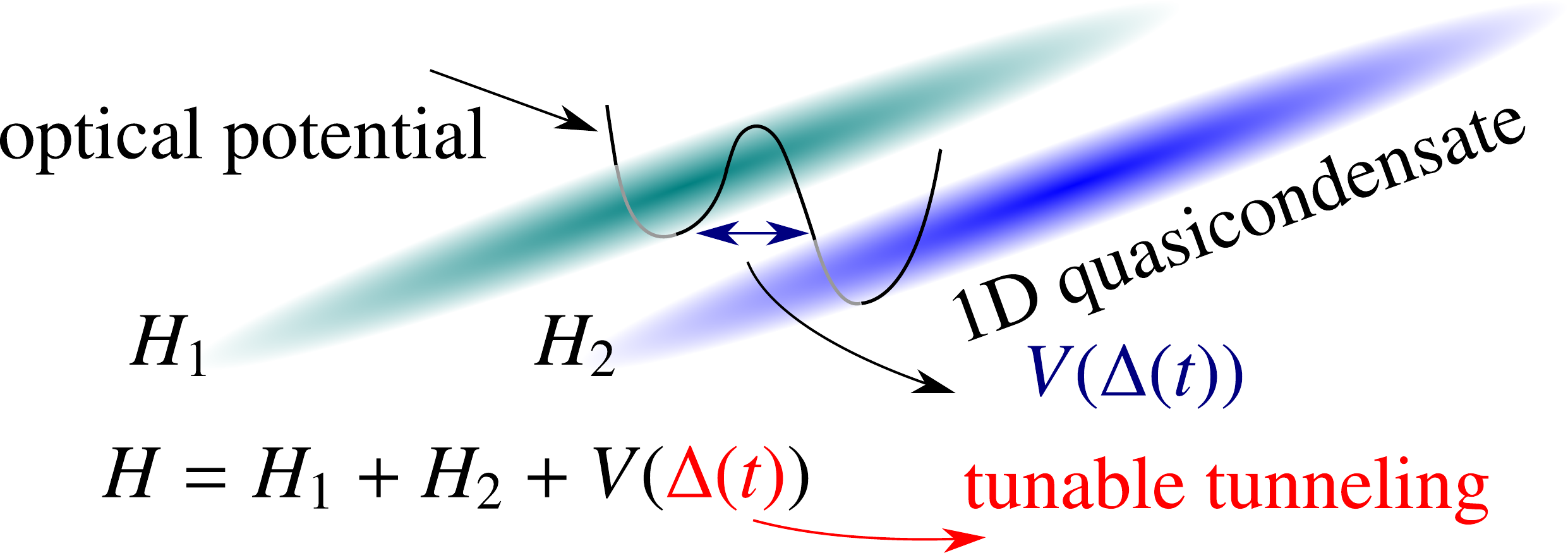}
 \caption{Two coupled one-dimensional condensates. The tunneling amplitude $\Delta(t)$ can be tuned by raising or lowering the potential barrier.}
 \label{fig:1}
\end{figure}

The outline of this paper is as follows. In Sec.~\ref{formulation}, we formulate our proposal generically, and briefly discuss our numerical simulated-annealing method. In Sec.~\ref{model}, we introduce the specific model of two coupled elongated condensates, and analyze its nonequilibrium quantum dynamics. Section~\ref{opt} presents numerical results on the optimal cooling protocols as well as the connection with the Pontryagin's maximum principle. We conclude the paper in Sec.~\ref{conc} with discussing a possible universal approach to cooling based on optimal control. In Appendix.~\ref{second}, we review the implications of the second law of thermodynamics for our cooling scheme. Finally, the equations of motion for our model system are derived in Appendix.~\ref{eom}.

\section{General formulation}\label{formulation}
Let us formulate our proposal generically. Consider a quantum system with Hamiltonian $H$, which is comprised of two coupled subsystems: $H=H_1+H_2+V$, where $H_i$ is the Hamiltonian of subsystem $i$ and $V$ is the coupling Hamiltonian. Generically, $V$ is a sum of certain local terms, with some coupling constants $\{\lambda\}$. We assume that (i) we have time-dependent control over the coupling constants $\{\lambda(t)\}$, i.e., within a range determined by the experimental constraints, we can tune them to any value as a function of time, (ii) for all initial $\{\lambda_0\}$, we can prepare the system at inverse temperature $\beta_0$ with the current state-of-the-art cooling methods, and, (iii) we have a fixed time $\tau$ to carry out a dynamical process (by tuning the Hamiltonian), during which the system undergoes quantum coherent unitary evolution.

Our scheme allows us to cool down at least one of the subsystems $H_i$ beyond the state-of-the-art temperature $1/\beta_0$ by performing unitary evolution on the thermally isolated compound system. Using two  coupled subsystems is essential for our scheme because according to Kelvin's statement of the second law of thermodynamics, the energy of a thermally isolated system, which is initially in thermal equilibrium, can not decrease by any cyclic dynamical process (see Appendix.~\ref{second}). However in the absence of many-body localization, energy can ``flow'' from one part of the system to another so it may be possible to devise dynamical processes which decrease the energy in one part of the system, say $H_1$. At the end of the process we set the coupling $V$ to zero and end up with a system described by Hamiltonian $H_1$, which has a lower excess energy than at thermal equilibrium at inverse temperature $\beta_0$.

Our goal is then to find an optimal protocol $\{\lambda(t)\}$ such that, at the end of the process ($t=\tau$), the energy of subsystem $1$, or some other custom-ordered cost function, is minimized. For a given protocol, the density matrix evolves as $\dot{\rho}(t)=i[H(\{\lambda(t)\}),\rho(t)]$, with initial conditions determined by the thermal state at $t=0$. Thus, $\rho(\tau)$, and, consequently, cost functions such as ${\cal E}_1(\tau)={\rm tr}\left[ H_1\rho(\tau)\right]$ are functionals of $\{\lambda(t)\}$, $0<t<\tau$. Notice that if we decouple the two subsystems at time $\tau$, the energy of subsystem $1$ does not change for subsequent times.

The key question addressed in this paper is how to minimize this functional of $\{\lambda(t)\}$. We find that i) the Pontryagin's maximum principle provides a deep understanding of the structure of such protocols, and ii) the simulated annealing method used in Ref.~\cite{Rahmani11} gives a simple and generic way for performing such optimization. In simulated annealing, we discretize time, approximate an arbitrary protocol by a piece-wise constant one, and perform direct (classical) Monte-Carlo (MC) simulations with kinetic moves consisting of small random displacements of randomly chosen pieces of the protocol~\cite{Rahmani11}. Such MC simulations explore the space of permissible protocols in an unbiased manner. At every step of the simulations, one needs to find the variation $\delta{\cal E}_1(\tau)$ of the cost function ${\cal E}_1(\tau)$ due to a small random variation of the piece-wise constant protocol and accept the variation with a probability proportional to $\exp \left[-\delta{\cal E}_1(\tau)/T_{\rm MC}\right]$, where $T_{\rm MC}$ is a fictitious temperature, which is gradually reduced to zero. This standard optimization method is often used in classical statistical mechanics. In this case, however, since ${\cal E}_1(\tau)$ depends on the entire protocol $\{\lambda(t)\}$, we have very nonlocal ``interactions'' between the degrees of freedom, i.e., the discrete values of a piece-wise constant $\{\lambda(t)\}$. In such simulations, we need to choose a large enough number of pieces (in the piece-wise constant protocol) so there is convergence in the optimal ${\cal E}_1(\tau)$.

\section{Dynamics of coupled condensates} \label{model}

Let us now discuss the specific system studied in this paper, i.e., a pair of coupled one-dimensional condensates of interacting atoms with Hamiltonian $H=H_1+H_2+V$, where
\begin{equation}
H_i=\frac{v_i}{2}\int d x\left[ \frac{\pi}{g_i}\Pi_i^2(x)+\frac{g_i}{\pi}\left(\partial_x \Phi_i(x)\right)^2\right].
\end{equation}
For $i=1,2$, $\Pi_i(x)$ is the conjugate momentum to bosonic field $\Phi_i(x)$, and the coupling term has a sine-Gordon form $V=-2 {\Delta \over a} \int dx \cos\left[\Phi_1(x)-\Phi_2(x)\right]$~\cite{Gritsev07}. Physically, $\Phi_i(x)$ and $\Pi_i(x)$ respectively represent the phase and the density fluctuations (with respect to a constant background density) of condensate $i$ at position $x$, and, $v_i$ and $g_i$ are respectively the sound velocity and the Luttinger parameter. As seen in Fig.~\ref{fig:1}, $\Delta \over a$ is an effective tunneling amplitude per length ($a$ is a microscopic length scale), which can be tuned by changing the height of the optical potential barrier~\cite{Schumm2005,Hofferberth2007,Hofferberth2008} (see also Refs. \cite{Onofrio2002,Onofrio2004}). This simple exactly solvable, and experimentally relevant, model provides a concrete demonstration of our method.

A comment on the dimensions of the quantities above is in order. We have set $\hbar$ to unity, and identified the units of time and inverse energy. Representing length and energy by $\ell$ and $\varepsilon$ respectively, the field $\Phi(x)$ is dimensionless, its conjugate momentum $\Pi(x)$ has dimension $\ell^{-1}$, and, $v$ and $\Delta$ respectively have dimension $\ell \varepsilon$ and $\varepsilon$. Let us now use the harmonic approximation (i.e., expand the cosine term around $\Phi_1(x)-\Phi_2(x)=0$ and keep the leading quadratic term). This approximation is justified (at least for the initial equilibrium state of two coupled condensates) in the limit of large Luttinger parameters where the cosine term is relevant. As we will check \textit{a posteriori}, although the differences $\Phi_1(x)-\Phi_2(x)$ typically increase by an optimal evolution designed to cool one of the condensates, for some range of parameters, one can keep them reasonably small during the evolution so that the harmonic approximation remains valid at all times.

We can then write the Hamiltonian in momentum space as a collection of harmonic oscillators:
\begin{eqnarray}\label{eq:Hamiltonian}
H&=&\sum_i\sum_{q>0}\left[{v_i \pi \over 4g_i}\left(\Pi^{\Re i}_q\right)^2+{v_i g_i \over  \pi}q^2\left(\Phi^{\Re i}_q\right)^2\right]\\
&+&\sum_{q>0}2{\Delta}\left(\Phi^{\Re 1}_q-\Phi^{\Re 2}_q\right)^2+\Re\leftrightarrow\Im,\nonumber
\end{eqnarray}
where $\Phi^{\Re (\Im) i}_q$ indicates the real (imaginary) part of $\Phi^i_q$, and $\Pi^{\Re (\Im) i}_q$ is the conjugate momentum to $\Phi^{\Re (\Im) i}_q$. Note that $\Phi_q$ and $\Pi_q$ respectively have dimension $\ell^{1/2}$ and $\ell^{-1/2}$. We have not included in Hamiltonian~\eqref{eq:Hamiltonian} the $q=0$ term $H_0={\pi \over 2 L} \sum_i{v_i \over g_i}\left(N^i-N^i_0\right)^2+ { \Delta L\over a}\left(\Phi^{ 1}_0-\Phi^{ 2}_0\right)^2$, which is responsible for changing the particle number $N^i$ of condensate $i=1,2$ ($\Phi^{i}_0$ is conjugate to $N^i$ and $N^i_0 \over L$ is the background density with $L$ representing the system size)~\cite{Haldane81}. Evolution with $H_0$ does not change the expectation values of $N_i$, but can change their fluctuations, which are neglected in this work. Note that to prepare number-squeezed states with optimal control, we need to work only with a single-mode Hamiltonian $H_0$~\cite{Grond09}.


For a given protocol $\Delta(t)$, each mode $q$ in the Hamiltonian~\eqref{eq:Hamiltonian} evolves independently. Although the modes do not interact in Eq.~\eqref{eq:Hamiltonian}, they all evolve with the same protocol $\Delta(t)$, which induces correlations between them. Therefore, we have a fundamentally many-mode problem even without (subleading) mode-coupling terms, which we have neglected. The first step, however, is to analyze the dynamics of a single mode $q$ consisting of just two coupled harmonic oscillators, namely, $H=H_1+H_2+H_{12}$, where $H_i=\frac{1}{2}(p_i^2/m_i+k_ix_i^2)\simeq\omega_ia_i^\dagger a_i$ and $H_{12}=\frac{\lambda}{2} (x_1-x_2)^2$ with
\begin{equation}\label{eq:parameters}
m_i={2g_i\over \pi v_i},\qquad k_i={2\over\pi}v_i g_i q^2,\qquad \lambda={4\Delta\over a}.
\end{equation}

Let us assume the initial thermal state is prepared at $\lambda=\lambda_0$. We then evolve the system with a time-dependent protocol $\lambda(t)$ (with the constraint $0<\lambda(t)<\lambda_{\rm max}=4 \Delta_{\rm max}/a$). For any $\lambda$, we can write the single-mode Hamiltonian as $H=\frac{1}{2}{ P}^{T}{ P}+\frac{1}{2}{ X}^{T}K(\lambda){ X}$, where ${P}^{T}=(\frac{p_1}{\sqrt{m_1}},\frac{p_2}{\sqrt{m_2}})$, ${X}^{T}=(\sqrt{m_1}x_1,\sqrt{m_2}x_2)$ and 
\[
 K(\lambda)=\left(\begin{array}{cc}
 (k_1+\lambda)/{m_1}&-{\lambda}/{\sqrt{m_1 m_2}}  \\ 
 -{\lambda}/{\sqrt{m_1 m_2}}&(k_2+\lambda)/{m_2}
\end{array}\right). 
\]
We can then diagonalize the above symmetric matrix as $K(\lambda)=Q(\lambda)\:\Omega(\lambda)\: Q^T(\lambda)$, where $Q(\lambda)$ is an orthonormal matrix of eigenvectors and $\Omega={\rm diag}( \bar{\omega}_1^2, \bar{\omega}_2^2)$, with $\bar{\omega}_i$ a normal-mode frequency.

For a system evolving with $\lambda(t)$, we can write the Heisenberg annihilation operator of oscillator $1$ (or $2$) in terms of the initial normal-mode operators as
\[
 a_1(t)=\sum_i \left[u_i(t) \bar{a}_i(\lambda_0)+v_i(t) \bar{a}^\dagger_i (\lambda_0)\right],
\]
where $u_i$ and $v_i$ are some complex coefficients, with initial conditions simply determined by $\lambda_0$ (see Appendix.~\ref{eom} for details). We can find the value of these coefficients at $t=\tau$ by integrating simple equations of motion (derived in Appendix.~\ref{eom}) from $t=0$ to $t=\tau$. It is helpful to define $\tilde{\lambda}(t)\equiv \lambda(\tau-t)$, which makes the equations of motion local in time. Finding the optimal $\tilde{\lambda}$ immediately yields the optimal $\lambda$. It is important for our discussion to emphasize that these equations of motion are \textit{linear} in $\tilde{\lambda}$.

The equations of motion for $u$ and $v$, together with their initial conditions, uniquely determine $a_1(\tau)$ as a functional of $\lambda(t)$. [Notice that the same equations with different initial conditions can be used to find $a_2(\tau)$ as well.] Our goal is to minimize an appropriate cost function, such as the excess energy of oscillator $1$, over all permissible controls $\lambda(t)$. For a single oscillator, the excess energy is proportional to the average number of excitations, which can be written as $\langle n_1(t)\rangle={\rm tr}\left[ a_1^\dagger(t)a_1(t) \: \rho_0 \right ]$, where the initial density matrix $\rho_0$ factorizes in terms of the normal-mode operators (see Appendix.~\ref{eom}). In terms of the dynamical variables $u_i(t)$ and $v_i(t)$, the trace above then simplifies to
\begin{equation}\label {eq:n}
 \langle n_1(t)\rangle=\sum_i |u_i(t)|^2 \bar{n}_i(0)+|v_i(t)|^2(1+ \bar{n}_i(0)),
\end{equation} 
where $\bar{n}_i(0)\equiv{\rm tr}\left[ \bar{a}_i(\lambda_0)^\dagger \bar{a}_i(\lambda_0) \: \rho_0 \right ]=\left({e^{ \beta_0 \bar{\omega}_i(\lambda_0) }-1}\right)^{-1}$. 

Note that exactly the same formulation describes the many-mode problem [Eq.~(\ref{eq:Hamiltonian})]. In this case, we have to multiply the number of dynamical variables by the number of modes. The equations of motion still hold for each mode, with parameters depending on $q$ as in Eq.~(\ref{eq:parameters}). Appropriate many-mode cost functions can be constructed from cost functions for individual modes. For example, we can simply add all $\langle n^{q}_1(t)\rangle$ to obtain the total number of excitations $\langle {\cal N}_1(t )\rangle$ in condensate $1$, or weight them by the mode frequency to find the total excess energy $\langle {\cal E}_1(t )\rangle$ in the condensate:
\begin{equation}\label {eq:cost}
 \langle {\cal N}_1(t )\rangle=2\sum_{0<q<\Lambda} \langle n^q_1(t )\rangle,
 \quad  \langle {\cal E}_1(t )\rangle=2 v_1\sum_{0<q<\Lambda} q\:\langle n^q_1(t )\rangle,
\end{equation}
where the factor of $2$ accounts for real and imaginary components of Hamiltonian~(\ref{eq:Hamiltonian}) and $\Lambda$ is a momentum cutoff. Additionally, we may also consider $\langle {\cal C}_1(t )\rangle=\sum_{0<q<\Lambda} \langle n^q_1(t )\rangle/q$, which is relevant for enhancing the fringe contrast of matter-wave interferometry experiments~\cite{Kitagawa10}.

\section{Cooling through optimal control}\label{opt}
The problem formulated thus far is a typical problem in optimal control theory applied to quantum dynamics~\cite{Hohenester07,Chen10, Doria11, Rahmani11,Hoffmann11,Caneva11,Choi2011, Choi2012}: we have a set of dynamical variables with given initial conditions ($u_i$ and $v_i$ in our case), which evolve with given equations of motion [Eqs.~(\ref{eq:u_eom}) and (\ref{eq:v_eom})] that depend on some admissible control parameter(s) [$0<\lambda(t)<\lambda_{\rm max}$]. The challenge is to find an admissible optimal control such that a given cost function of the dynamical variables [Eq.~(\ref{eq:cost}) in our case] is minimized at a given time $\tau$.

Let us now turn to the main questions of this work: What do the optimal $\lambda(t)$  protocols look like? How can we find them? How much can they cool a system? Using Pontryagin's maximum principle, we argue that optimal protocols are \textit{bang-bang}, i.e., $\lambda(t)$ is either zero or equal to $\lambda_{\rm max}$  at any given time. As mentioned earlier, we demonstrate that a direct simulated-annealing calculation can yield these optimal protocols. We also find that, depending on the parameters of the problem, it is possible to significantly cool down one of the condensates.

Let us now briefly review Pontryagin's maximum principle. Consider a set of dynamical variables $\{x(t)\}$ that satisfy the equations of motion
$\dot{{x}}_j=f_j(\{{x},\alpha\})$, with ${x}_j(0)={x}_j^0$, for a set of admissible controls $\{\alpha(t)\}$. The goal is to maximize a payoff function $g(\{{x(\tau)}\})$ over all such $\{\alpha(t)\}$. The key to Pontryagin's maximum principle is the following \textit{optimal-control Hamiltonian}:
\begin{equation}\label{eq:och}
{\mathscr H}(\{x,p,\alpha\})=\sum_j  p_j(t)\:f_j(\{x,\alpha\}),
\end{equation}
where $p_j(t)$ is a ``momentum'' conjugate to $x_j(t)$.
The Pontryagin's theorem states that for the optimal control $\{\alpha^*(t)\}$, and the corresponding $\{x^*,p^*\}$, we have
\begin{equation}\label{eq:pontryagin}
{\mathscr H}^*\equiv{\mathscr H}(\{x^*,p^*,\alpha^*\})=\max_{\{\alpha\}}{\mathscr H}(\{x^*,p^*,\alpha\}),
\end {equation}
where $x$ and $p$ satisfy $\dot{x}^*_j=\frac{\partial {\mathscr H}^*}{\partial{p^*_j}}$ and $\dot{p}^*_j=-\frac{\partial {\mathscr H}^*}{\partial{x^*_j}}$
with boundary conditions ${x}^*_j(0)={x}_j^0$ and $p^*_j(\tau)={\partial \over  \partial x_j} g(\{x^*(\tau)\})$.
It is now easy to observe that since, for all modes $q$, the equations of motion for $u_i$ and $v_i$ are linear in $\lambda(t)$, the optimal-control Hamiltonian [Eq.~(\ref{eq:och})] is also a linear function of $\lambda(t)$ in our case. We then immediately deduce from Eq.~(\ref{eq:pontryagin}) that, unless ${\mathscr H}^*$ identically vanishes over a finite time interval, the control $\lambda(t)$ can only take two values, namely, zero and $\lambda_{\max}$. This is a generic feature of a Hamiltonian that is linear in the tunable coupling constants.

The Pontryagin equations are not easy to solve numerically for many modes. We thus use our direct MC method, without utilizing any assumptions regarding the bang-bang nature of the protocol. The simulations consist of varying a randomly chosen $\lambda_i$ (of the discretized protocol) by a small random amount, and accepting or rejecting the variation based on the change in the cost function. As found in Ref.~\cite{Rahmani11}, such simulations converge very well in the number of discretization points. For the exactly solvable model studied here, the cost function for an arbitrary protocol can be determined very efficiently: we define new dynamical variables $\phi_j \equiv \sqrt{\bar{\omega}_{j}(\lambda_0)}\left(u_j+v_j\right)$ and $\theta_j \equiv \left(u_j-v_j\right)/\sqrt{\bar{\omega}_{j}(\lambda_0)}$, which satisfy $|\dot{\phi}\rangle=-i\bar{K}(\tilde{\lambda})|\theta\rangle$ and $|\dot{\theta}\rangle=-i|\phi\rangle$ in matrix notation (this change of variables allows us to diagonalize $2\times 2$ matrices instead of $4 \times 4$ ones). By solving the above equations in terms of the eigenvalues and eigenvectors of the  $2\times 2$ matrix $\bar{K}(\tilde{\lambda}_m)=\bar{K}({\lambda}_{n-m+1})$, we can then write simple recursion relation for $|\phi\rangle$ and $|\theta\rangle$, which yield their values at time $\tau$ after $n$ iterations.

\begin{figure}
 \centering
 \includegraphics[width =7.8cm]{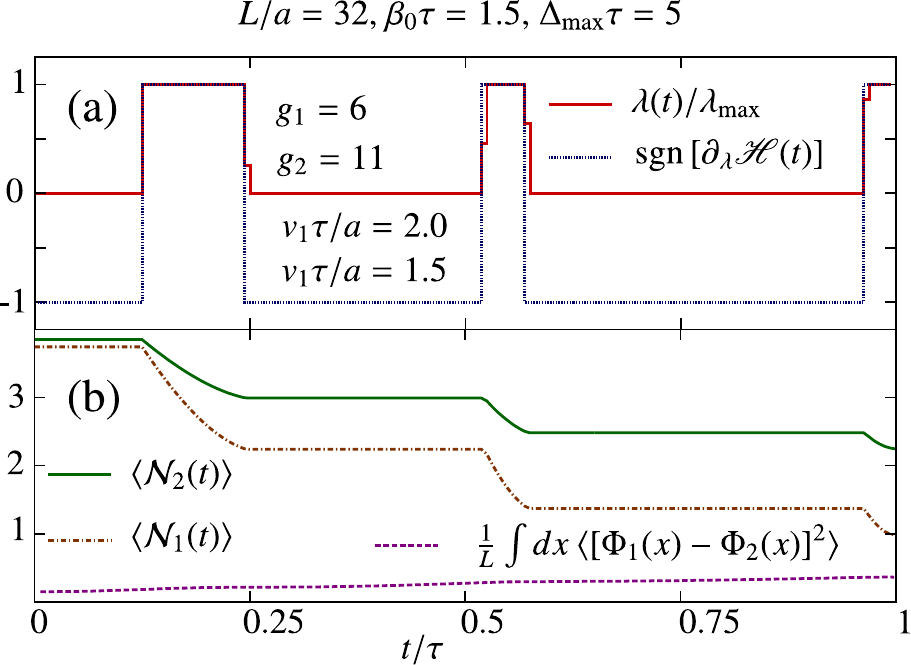} 
 \caption{(a) Typical protocols obtained with unbiased simulated annealing, and the derivative of the optimal control Hamiltonian with respect to control $\lambda$, whose sign determines the protocol. The simulations converge to bang-bang protocols in excellent agreement with ${\rm sgn}\left[\partial_\lambda \mathscr H\right]$. (b) Reduction of $\langle{\cal N}_i(t)\rangle$ due to evolution with the two optimal protocols above. Cooling condensate 1, may also cool down the other condensate for free. For large Luttinger parameters, the argument of the cosine term remains much smaller than one during the evolution, i.e., the harmonic approximation remains valid. }
\label{fig:2}
\end{figure}

One comment is in order before proceeding. In addition to optimizing over $\lambda(t)$, we have the freedom to choose the initial $\lambda_0$ anywhere between zero and $\lambda_{\rm max}$. There is a rigorous lower bound on $\langle n_1\rangle$ [Eq.~(\ref{eq:n})]: $\langle n_1(t)\rangle \geqslant n_{\rm min}\equiv \min (\langle \bar{n}_1(0)\rangle,\langle \bar{n}_2(0)\rangle)$. We can show that $n_{\rm min}$ is a decreasing function of $\lambda_0$. This suggests that it may be advantageous to set $\lambda_0=\lambda_{\rm max}$. Although the actual cost functions we are able to reach by our minimization procedure are typically much larger than this lower bound, by trying several values of $\lambda_0$ in our numerics, we have found that the best cooling is in fact achieved for $\lambda_0=\lambda_{\rm max}$.
In Fig.~\ref{fig:2}(a), we show a typical protocol obtained by MC simulations. We converge to a bang-bang protocol by an unbiased simulation, which samples all the intermediate values of $\lambda$, and, \textit{a priori}, does not assume anything about the shape of the protocol. Surprisingly, minimizing ${\cal N}_1$, ${\cal E}_1$, or ${\cal C}_1$ leads to very similar, albeit nonidentical, protocols (here we show the protocol obtained by minimizing ${\cal N}_1$).

To further check the consistency with the Pontryagin's theorem, we also computed the derivative $\partial_\lambda \mathscr H$, the sign of which determines $\lambda(t)$ through Eq.~(\ref{eq:pontryagin}), for cost function $\langle{\cal N}_1(\tau)\rangle$.
To construct $\mathscr H$, we need to treat the real and imaginary parts of $\phi_j$ and $\theta_j$ as separate dynamical variables with their own conjugate momenta. We can then construct a complex variable ${\pi}^{\phi }_j$, whose real (imaginary) part is the conjugate momentum to the real (imaginary) part of ${\phi }_j$, and similarly for ${\theta}_j$. For each $q$, we then have $|\dot{\pi}^{\phi }\rangle=-i |\pi^{\theta }\rangle$ and $|\dot{\pi}^{\theta }\rangle=-i \bar{K} (\tilde{\lambda})|\pi^{\phi }\rangle$. The boundary conditions at $t=\tau$ depend on the cost function [see the boundary conditions below Eq.~\eqref{eq:pontryagin}] and, for $\langle{\cal N}_1(\tau)\rangle$, can be written as $\pi^{\phi }_{i}(\tau)=\theta_{i}(\tau)-\left(2\langle\bar{n}_i(0)\rangle+1 \right){1 \over \bar{\omega}_i(0)}$ and $\pi^{\theta }_{i}(\tau)=\phi_{i}(\tau)-\left(2\langle\bar{n}_i(0)\rangle+1 \right) \bar{\omega}_i(0)
\:\theta_{i}(\tau)$.
Given a protocol $\lambda(t)$, we can solve for $\phi$ and $\theta$ forward in time, construct $\pi^{\phi }(\tau)$ and $\pi^{\theta }(\tau)$ from the boundary conditions above, solve for  $\pi^{\phi }$ and $\pi^{\theta }$ backward in time and finally construct $\partial_{\tilde{\lambda}} \mathscr H=\sum_{q}
\langle\pi^{\phi }_{q}|\partial_{\tilde{\lambda}}\bar{K}_q(\tilde{\lambda})|\theta_q\rangle $, which immediately yields $\partial_{\lambda} \mathscr H$. The results are shown in Fig.~\ref{fig:2}(a), and show excellent agreement with the simulations.

In Fig.~\ref{fig:2}(b), we show how the cost function $\langle {\cal N}_1(t)\rangle$ changes when evolving with the optimal protocol. An interesting feature of the evolution is that ${d\langle {\cal N}_1(t)\rangle\over dt}\neq 0$ just before quenching to $\lambda(t)=0$. Keeping the subsystems coupled would do a better job in reducing $\langle {\cal N}_1(t)\rangle$ locally (in time) but would not lead to global optimization in total time $\tau$. We can also check the harmonic approximation \textit{a posteriori} by computing ${1 \over L}\int dx\: \langle\left[\Phi_1(x)-\Phi_2(x)\right]^2\rangle$. We find that as long as the approximation is valid initially, and $\lambda_{\rm max}$ is large enough, this quantity remains smaller than one and the harmonic approximation holds throughout the evolution (if the system is not too long, the spatial fluctuations of $\left[\Phi_1(x)-\Phi_2(x)\right]^2$ are small). Also as the number of excitations in both condensates decreases monotonically, the Luttinger description remains valid and the results do not depend on the cutoff.

Interestingly, the optimal protocol designed for reducing the energy of condensates $1$ turns out to also cool down condensate $2$. This is not a violation of the second law of thermodynamics as our process is not cyclic: we start from two coupled condensates with $H=H_1+H_2+V$, and end up with two decoupled ones with $H=H_1+H_2$. The process only reduces the expectation value $\langle H_1+H_2\rangle$, while $\langle H_1+H_2+V\rangle$, which corresponds to the initial Hamiltonian, actually increases.
\section{Discussion and Conclusions}\label{conc}

The effective cooling described here is an out-of-equilibrium reduction of the excess energy, and does not imply thermal equilibrium. If the low-energy system equilibrates afterwards, however, it will have a lower temperature. To directly bring the system close to thermal equilibrium, one can instead minimize the trace distance between the density matrix and thermal density matrices at varying target temperatures, and find a balance between a small trace distance and a low target temperature. We do not pursue this approach here because, under realistic circumstances, each decoupled condensate is expected to eventually decohere, and reach an effective temperature determined by its excess energy~\cite{Burkov07}.

The performance of our optimal protocols depends on several (dimensionless) parameters, including the two Luttinger parameters, $\beta_0 \tau$, $v_i \tau /a$, and $L/a$. However, $\Delta_{\rm max}\tau$ and the ratio $v_2/v_1$ seem to have the most pronounced effect on the performance (measured by the ratio of the achieved energy to equilibrium energy at $\beta_0$). Typically, the energy can be reduced by a factor of $3$ to $5$ when $v_2/v_1$ is of order unity. For a highly asymmetric system with $v_2/v_1=100$, we achieved an energy reduction by a factor of $40$ with a system size of $L/a=64$ and other dimensionless parameters of order unity.

Let us now comment on a possible extension of our scheme to arbitrary systems. To perform our MC simulations, we need to be able to efficiently compute desired cost functions for any allowed protocol (which is the case for our system in the harmonic approximation). Cooling down more complicated systems such as the two-dimensional fermionic Hubbard model (or even our system in regimes where the full sine-Gordon term is needed) is of considerable interest for quantum simulations. The generality of our MC method, however, raises an intriguing possibility for a \textit{universal} approach: if one can automate the processes of system initialization (initial cooling), unitary evolution (with a tailored protocol), and measurement of the figure of merit (e.g., energy), then the \textit{system itself} can be used to perform such MC simulations. The cost function can be measured (instead of computed), and then fed into the MC algorithm. This would provide a powerful means of preparing desired states in arbitrary systems, and may open the door to the quantum simulation of unsolved condensed-matter models. Such integration of experiment and simulation has in fact been applied to the control of chemical reactions~\cite{Assion98}, and more recently to some aspects of cold atom experiments~\cite{Rohringer08}. In the absence of many-body localization, the energy can generically flow in quantum systems and we expect our scheme for cooling through optimal control, which we have explicitly demonstrated for a solvable model, to work for arbitrary clean systems.

In summary, we demonstrated that nonadiabatic optimal control of quantum evolution can be used to push the boundaries of atomic cooling. Contrary to the conventional association of nonadiabatic effects with heating, we showed that breaking away from the adiabatic limit, in a controlled way, can in fact help cool down quantum systems by directing the flow of energy. We applied this idea to a system of two coupled elongated condensates. Through simple and direct MC simulations, we found optimal protocols which agree with theoretical predictions based on Pontryagin's maximum principle, and are effective in reducing the excess energy. Such MC simulations can be potentially performed by the system itself giving access to a universal cooling scheme.

\appendix
\section{Unitary evolution and the second law}\label{second}
In this appendix we review a modern quantum derivation of the Kelvin's statement of the second law of thermodynamics~\cite{Thirring}. Explicitly, we show that the expectation value of the energy of a thermally isolated quantum system, which is initially in thermal equilibrium, cannot decrease if its Hamiltonian changes in a cyclic (but otherwise arbitrary) manner.

Denoting the eigenvalues of the initial Hamiltonian by $\epsilon_i$, and the elements in the energy basis of the initial density matrix by $\rho_i \propto e^{-\beta_0 \epsilon_i}$, the initial average energy is given by ${\cal E}(0)=\sum_i \epsilon_i\rho_i $. If the system undergoes unitary evolution with an arbitrary evolution operator $U$ (with matrix elements $U_{ij}$ in the same energy basis), the final energy for a cyclic process is given by ${\cal E}(\tau)=\sum_{ij} \epsilon_i W_{ij}\rho_j$, where $W_{ij}\equiv U^{\dagger}_{ij}U_{ji}$.
The matrix $W$ is \textit{doubly stochastic} (i.e., $\sum_iW_{ij}=\sum_jW_{ij}=1$). Now according to von Neumann-Birkhoff theorem (see, e.g., Ref.~\cite{Graham95}), any such matrix can be written as a convex combination of permutation matrices, i.e., $W=\sum_k c_k P_k$, where $P_k$ is a permutation matrix and $\sum_k c_k=1$ for positive scalar $c_k$. Therefore
${\cal E}(\tau)=\sum_{ijk} c_k\epsilon_i P^k_{ij}\rho_j $.
In the initial thermal matrix $\epsilon_i> \epsilon_j$ implies $\rho_i< \rho_j$, so for any permutation of the weights $\rho_i$, $\sum_{ij} \epsilon_i P^k_{ij}\rho_j \geqslant \sum_{i} \epsilon_i \rho_i$. Since $\sum_k c_k=1$, we immediately obtain
${\cal E}(\tau)\geqslant {\cal E}(0)$.

\section{Equations of motion}\label{eom}
Here we present some details regarding the initial conditions and the equations of motion of our system.
 In terms of $\bar{\omega}_{1,2}(\lambda_0)$, the initial density matrix of a system of two coupled oscillators is given by
\[\rho_0=\frac{1}{\cal Z}e^{-\beta_0 \bar{\omega}_{1}(\lambda_0) \bar{a}_1^\dagger(\lambda_0) \bar{a}_1(\lambda_0)}e^{-\beta_0 \bar{\omega}_{2}(\lambda_0) \bar{a}_2^\dagger (\lambda_0)\bar{a}_2(\lambda_0)},
\]
where $\bar{a}_j(\lambda_0)$  is the annihilation operator for normal-mode $j=1,2$, which can be written in terms of the annihilation operators $a_i$ of oscillators $i=1,2$ as
\begin{eqnarray}\label{eq:initial}
\bar{a}_{j}&=&{1 \over 2}\sum_k Q_{kj}
\left(
{\cal F}_{jk}\:a_k+ 
{\cal G}_{jk}\:a^\dagger_k
 \right),\\
{\cal F}_{jk}&\equiv& \sqrt{\frac{\bar{\omega}_j}{\omega_k}}+ 
\sqrt{\frac{\omega_k}{\bar{\omega}_j}},\quad 
{\cal G}_{jk}\equiv\sqrt{\frac{\bar{\omega}_j}{\omega_k}}- \sqrt{\frac{\omega_k}{\bar{\omega}_j}} .
\end{eqnarray}
The  initial conditions (at $t=0$ ) for  the coefficients $u_i$ and $v_i$ are obtained by inverting Eq.~(\ref{eq:initial}), i.e., $a_j={1 \over 2}\sum_k Q_{jk}
\left(
{\cal F}_{kj}\:\bar{a}_k-
{\cal G}_{kj}\:\bar{a}^\dagger_k
\right)$. Note that these coefficients must satisfy the constraint
$
|u_1|^2+|u_2|^2-|v_1|^2-|v_2|^2=1
$
to preserve the commutation relations.
 
 To compute $u_i(\tau)$ and $v_i(\tau)$, it is convenient to consider a piece-wise constant protocol determined by a sequence $(\lambda_i,t_i)$, for $i=1 \dots n$, so that $$a_1(\tau)=e^{i H(\lambda_1)t_1}\dots e^{i H(\lambda_n)t_n} a_1(0)e^{-i H(\lambda_n)t_n}\dots e^{-i H(\lambda_1)t_1}.$$
Using the commutation relation 
\[
 [\bar{a}_i,H(\lambda)]=\frac{\bar{\omega}_i}{2}(\bar{a}_i-\bar{a}_i^\dagger)+{1\over 2}\sum_j{\bar{K}_{ij}(\lambda)\over\sqrt{\bar{\omega}_i\bar{\omega}_j}} (\bar{a}_j+\bar{a}_j^\dagger),
\]
with $\bar{K}(\lambda)=Q^T(\lambda_0) K(\lambda) Q(\lambda_0)$, we then find that these coefficients at $t=\tau$ are obtained by integrating the following equations of motion from $t=0$ to $t=\tau$: 
\begin{eqnarray}\label{eq:u_eom}
 \dot{u}_j &=& {1 \over 2 i}\left[\sum_k (u_k-v_k){\bar{K}_{jk}(\tau-t)\over\sqrt{\bar{\omega}_j\bar{\omega}_k}} +(u_j+v_j){\bar{\omega}_j}\right],   \\
  \dot{v}_j&=& {1 \over 2 i} \left[ \sum_k (u_k-v_k){\bar{K}_{jk}(\tau-t)\over\sqrt{\bar{\omega}_j\bar{\omega}_k}} -(u_j+v_j)\bar{\omega}_j\right],\label{eq:v_eom}
\end{eqnarray}
where all normal-mode frequencies $\bar{\omega}$ are calculated at $\lambda=\lambda_0$. Notice that the equations above depend on the final time $\tau$, and should not be used for computing $u_i$ and $v_i$ at $t\neq \tau$. When working at fixed $\tau$, it is helpful to define $\tilde{\lambda}(t)\equiv \lambda(\tau-t)$, which makes the equations local in time. 

\acknowledgements
We are grateful to A. Polkovnikov, J. Schmiedmayer, and M. Zwierlein for helpful discussions. This work was supported in part by the U.S. DOE under LANL/LDRD program (A.R.), DOE Grant No. DE-FG02-06ER46316 (C.C.), Harvard-MIT CUA (E.D. and T.K.), NSF Grant No. DMR-07-05472 (E.D. and T.K.), AFOSR Quantum Simulation MURI (E.D. and T.K.), the ARO-MURI on Atomtronics (E.D. and T.K.), and ARO MURI Quism program (E.D. and T.K.).

\bibliography{cooling}{}

\end{document}